\documentclass{icrc29}
\usepackage{graphicx,amssymb,amsmath,times}
\begin{document}
\title[Search for massive rare particles...]{Search for massive rare particles with the SLIM experiment}
\author[S. Balestra et al.] {S.Balestra $^1$, S. Cecchini$^{1,7}$, F. Fabbri $^1$, G. Giacomelli$^1$, A. Kumar $^{1,6}$, S. Manzoor $^{1,4}$
\newauthor
J. McDonald $^3$, E. Medinaceli $^1$, J. Nogales $^5$, L. Patrizii $^1$, J. Pinfold $^3$, V. Popa $^1$, 
\newauthor
I. Qureshi $^4$, O. Saavedra $^2$, G. Sher $^4$, M. Shahzad $^4$, M. Spurio $^1$, R. Ticona $^5$, V. Togo $^1$,
\newauthor 
 A. Velarde $^5$, A. Zanini $^2$ 
         \\
        (1) Dip. Fisica dell'Universita di Bologna and INFN, 40127 Bologna, Italy \\ 
        (2) Dip. Fisica Sperimentale e Generale dell'Universita di Torino and INFN, 10125 Torino, Italy\\
        (3) Centre for Subatomic Research, Univ. of Alberta, Edmonton, Alberta T6G 2N4, Canada\\
        (4) PRD, PINSTECH, P.O. Nilore, Islamabad, Pakistan\\
        (5) Laboratorio de Fisica Cosmica de Chacaltaya, UMSA, La Paz, Bolivia\\
        (6) Dept. of Physics, Sant Longowal Institute of Eng. \& Tech., Longowal, 148 106 India\\
        (7) INAF/IASF Sez. Bologna, 40129 Bologna, Italy\\
        }
\presenter{Presenter: S. Cecchini (cecchini@bo.infn.it), \  
ita-cecchini-S-abs2-he23-oral}

\maketitle

\begin{abstract}

The search for magnetic monopoles in the cosmic radiation remains one of the main aims of non-accelerator particle astrophysics. Experiments at high altitude allow lower mass thresholds with respect to detectors at sea level or underground. The SLIM experiment is a large array of nuclear track detectors at the Chacaltaya High Altitude Laboratory (5290 m a.s.l.). The results from the analysis of 171 m$^2$ exposed for more than 3.5 y are here reported. The completion of the analysis of the whole detector will allow to set the lowest flux upper limit for Magnetic Monopoles in the mass range 10$^5$ - 10$^{12}$ GeV. The experiment is also sensitive to SQM nuggets and Q-balls, which are possible Dark Matter candidates. 

\end{abstract}

\section{Introduction}
Cosmic rays are the most likely "site" to search for massive Magnetic Monopoles (MMs), since accelerator energies are insufficient to produce them. Grand Unification Theories (GUT) of strong and electroweak interactions at the high energy scale M$_G$ $\sim$  10$^{14}$ $\div $ 10$^{15}$ GeV predict the existence of MMs, produced in the early Universe at the end of the GUT epoch, with extraordinarily large masses M$_{MM}$ $\sim$ 10$^{16}$ $\div $ 10$^{17}$ GeV. GUT poles should be characterized by low velocity and relatively large energy losses \cite{MMs}. At present the MACRO experiment has set the best limit on GUT MMs for 4 10$^{-5}$ $<\beta=v$/c$< $ 0.5 \cite{MACRO}. Intermediate Mass Monopoles (IMMs) [10$^{7}$ $\div $ 10$^{12}$ GeV] with magnetic charge g = 2 g$_{D}$ could also be present in the cosmic radiation; they may have been produced in later phase transitions in the early Universe \cite{IMMs}. The recent interest in relatively low mass MMs is also connected with the possibility that they could yield the highest energy cosmic rays \cite{UHECR}. These particle, in fact, can have relativistic velocities since they could be accelerated to high velocities in one coherent domain of the galactic magnetic field. In this case one would have to look for fast ($\beta>$0.1) heavily ionizing MMs.

Besides MMs, other massive particles have been hypothesized to exist in the cosmic radiation and possibly to be components of the galactic cold dark matter: nuggets of Strange Quark Matter (SQM)- called nuclearites when neutralized by captured electrons - and Q-balls.

SQM should consist of aggregates of u, d and s quarks in approximately equal proportions \cite{nuclr}. It has been proposed that SQM may be the ground state of QCD. SQM nuggets should be stable for all baryon numbers in the range between ordinary heavy nuclei and neutron stars (A$\sim$ 10$^{57}$).
They could have been produced in the early Universe or in violent astrophysical processes, and may be present in the cosmic radiation. Nuclearite interaction with matter could depend on their mass and size. In \cite{SLIM05/5} different mechanisms of energy loss and propagation in relation to their detectability with the SLIM apparatus (described below) are considered. In the absence of any candidate, SLIM will be able to rule out some of the hypothesized propagation mechanisms.

Q-balls are super-symmetric coherent states of squarks, sleptons and Higgs fields, predicted by minimal super-symmetric generalizations of the Standard model \cite{qballs} they could have been produced in the early Universe. Charged Q-balls should interact with matter in ways not too dissimilar from those of nuclearites. 
 
 Fig. 1 shows the experimentally accessible region (minimal velocity at the top of the atmosphere versus the nuclearite mass) for arrays of Nuclear Track Detectors (NTDs) located at different altitudes. A similar plot for MMs can be found in \cite{ICRC03}.

\begin{figure}[h]
\vspace{-1.1in}
\begin{minipage}[c]{.35\textwidth}
 \centering
 \caption{\label {fig1} Accessible regions in the plane (mass, $\beta$) for  nuclearites coming from above for experiments at high altitudes and underground.}
 \end{minipage}%
 \hspace{.05\textwidth}
 \begin{minipage}[c]{.55\textwidth}
 \centering
 \includegraphics*[width=1.15\textwidth,angle=0,clip]{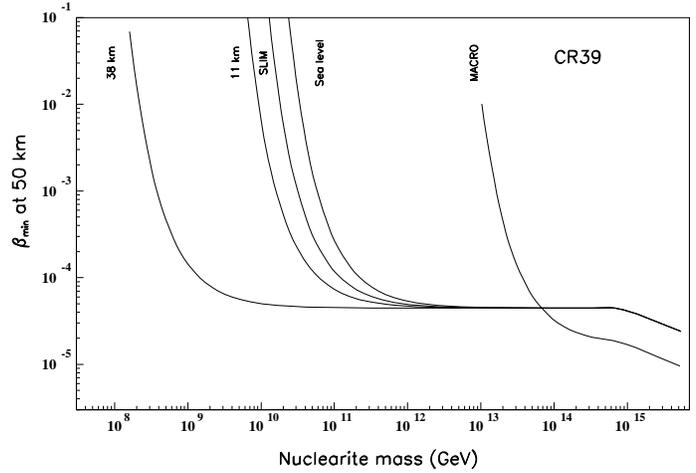}
 \end{minipage}%
 \end{figure}
In the followings, after a brief description of the apparatus we present the calibration, the analysis procedures and the results from 171 m$^2$ of the Chacaltaya's detectors after 3.5 years of exposure.

\section{The SLIM apparatus}
The SLIM (Search for LIght magnetic Monopole) experiment, based on 440 m$^2$ of NTDs, has been deployed at the Chacaltaya High Altitude Laboratory (Bolivia, 5290 m a.s.l.) since 2001. Another 100 m$^2$ of NTDs have been installed at Koksil (Pakistan, 4600 m a.s.l.) since 2003. 
The complete description of the apparatus is given in \cite{ICRC03}. Here we recall only a few features.
The detector modules have been exposed under the roof of the  Chacaltaya Laboratory at a height of 4 m from ground. The air temperatures were recorded every day at 8:00, 12:00 and 18:00 local time together with the minimum and maximum values. The observed range of temperatures (0 - 25$^\circ$ C) allow us to conclude that no significant time variations have occurred in the detector response  of the CR39 and Makrofol. Moreover the aluminized plastic bags in which the NTDs were sealed did not show any appreciable leakage of air (oxygen) after 3.5 years of exposure.

We have reported in \cite{ICRC03} the measurements of radon activity and the flux of cosmic ray neutrons. The early measurements are in agreement with more recent measurements at the same location \cite{neutron}.

\section{Calibrations}
Extensive test studies were made in order to improve the etching procedures of CR39 and Makrofol NTDs, improve the scanning and analysis procedures and speed, and keep a good scan efficiency. 
"Strong" and "soft" etching conditions have been defined for CR39 and Makrofol NTDs \cite{NTDsM},\cite{NTDsL}.
 
CR39 "strong" etching conditions -  8N KOH + 1.25\% Ethyl alcohol at 77$^\circ$ C for 30 hours. The strong etching is used for the first CR39 sheet in each module, in order to produce large tracks, easier to detect during scanning. 

CR39 "soft" etching conditions  -  6N NaOH + 1\% Ethyl alcohol at 70$^\circ$ C for 40 hours. The soft etching is applied to the other CR39 layers in a module, if a candidate track is found in the first layer. It allows more reliable measurements of the restricted energy loss (REL) and of the direction of the incident particle.

Makrofol layers are etched in 6N KOH + Ethyl alcohol (20\% by volume), at 50$^\circ$ C.

The detectors have been calibrated using 1 A GeV $^{26+}$Fe from the BNL AGS, with 158 A GeV $^{49+}$In and 30 A GeV $^{82+}$Pb beams at the CERN SPS. For "soft" etching conditions the threshold in CR39 is at REL $\sim$ 50 MeV cm$^2$ g$^{-1}$; for strong etching the threshold is at  REL $\sim$ 200 MeV cm$^2$ g$^{-1}$. 
The Makrofol polycarbonate has a higher threshold (REL$ \sim$ 2.5 GeV cm$^2$ g$^{-1}$). More details on calibration can be found in \cite{NTDsL}. 
It results that the CR39 allows the detection of IMMs  with two units Dirac charge in the whole $\beta$-range of 4 10$^{-5}$ $<$ $\beta$ $<$ 1. 
The Makrofol is useful for the detection of fast MMs. Nuclearites with  $\beta \sim$ 10$^{-3}$ can be detected by both CR39 and Makrofol \cite{NTDsM}. 

	The acceptance of the SLIM detector for MMs computed for the "soft" etching conditions is plotted in Fig. 2 for g = g$_{D}$, 2g$_{D}$, 3g$_{D}$ and dyons.
	
\begin{figure}[h]
\begin{minipage}[c]{.45\textwidth}
 \centering
 \caption{\label {fig2} Acceptance for "soft" etching of the SLIM apparatus for MMs with g = g$_{D}$, 2g$_{D}$, 3g$_{D}$ and for dyons (M+p , g = g$_{D}$)}
 \end{minipage}%
 \hspace{.05\textwidth}
 \begin{minipage}[c]{.55\textwidth}
 \vspace{0.1in}
 \centering
 \includegraphics*[width=0.8\textwidth,angle=0,clip]{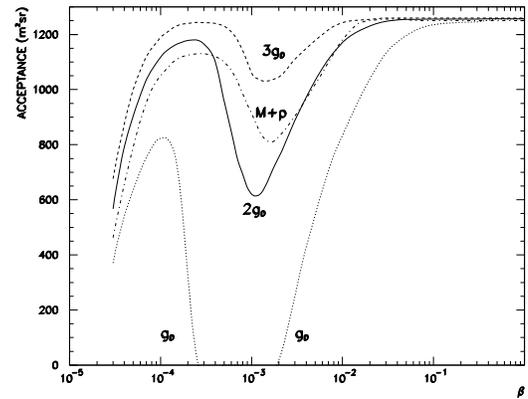}
 \end{minipage}%
 \end{figure}

\vspace{-0.1in} 
\section{Analysis}
	The analysis of a SLIM module begins by etching the top CR39 sheet using  "strong" conditions in order to quickly reduce its thickness from 1.4 mm to $\sim$0.6 mm. Since MMs, nuclearites and Q-balls have a constant REL through the stack, the signal looked for is a hole or a biconical track with the two base-cone areas equal within the experimental uncertainties. After the strong etching the sheets are scanned with a stereo microscope searching for a signal at low magnification (Field of View $\sim$ 1cm$^2$).  
Possible candidates are marked and further analysed under a high magnification microscope. The size of surface tracks is measured on both sides of the sheet.  We require the two values to be equal within 3 times the standard deviation of their difference. Finally a track is defined as  a "candidate" if the REL and the incidence angles on the front and back sides are equal to within 15\%. For each candidate the azimuth angle and its position referred to the fiducial marks are determined.  
	To confirm the candidate track the bottom CR39 layer is then etched in the "soft" conditions; an accurate scan under an optical microscope with high magnification is performed  in a region of about 0.5 mm around the expected candidate  position. If a two-fold coincidence is found the middle layer of the CR39 (and, in case of a high Z candidate, the Makrofol layer) is analyzed with soft conditions.

	Up to now no two-fold coincidence has been found, that is no magnetic monopole, nuclearite or Q-ball candidate was detected.

\section{Results and Conclusions}
 $\sim$171 m$^2$ of CR39 have been etched and analysed, with an average exposure time of 3.5 years. No candidate passed the searching criteria: the 90\% C.L. flux upper limits for fast ($\beta>$0.1) IMM's, nuclearites and Q-balls of any speed, all coming from above, are at the level of 3.9 10$^{-15}$ cm$^{-2}$ sr$^{-1}$ s$^{-1}$.
 
By the end of 2006 the 440 m$^2$ analysis will be completed and the experiment will reach a sensitivity of 10$^{-15}$ cm$^{-2}$ sr$^{-1}$ s$^{-1}$ for $\beta \geq$10$^{-2}$ and IMMs with 10$^7$ $<$ M$_{IMM}$ $<$ 10$^{13}$ GeV; the same sensitivity should be reached also for nuclearites and Q-balls with galactic velocities.
 Moreover this search will benefit from the analysis of further 100 m$^2$ of NTDs  installed at Koksil (Pakistan).

\section{Acknowledgements}
We acknowledge the collaboration of E. Bottazzi, L. Degli Esposti, R. Giacomelli and C. Valieri of INFN, Sez. di Bologna and the technical staff of the Chacaltaya High Altitude Laboratory. We thank INFN and ICTP for providing grants for non-italian citizens.

\end{document}